\documentstyle[epsf,prd,aps,epsfig]{revtex}

\newcommand{\bq}{\begin{equation}}
\newcommand{\eq}{\end{equation}}
\newcommand{\bqn}{\begin{eqnarray}}
\newcommand{\eqn}{\end{eqnarray}}
\newcommand{\nb}{\nonumber}
\newcommand{\lb}{\label}

\begin{document}

\twocolumn[\hsize\textwidth\columnwidth\hsize\csname
@twocolumnfalse\endcsname
 
\title{Collapsing and Expanding  Cylindrically Symmetric  Fields with
Ligh-tlike Wave-Fronts in General Relativity}  
\author{Paulo R.C.T.
Pereira\thanks{Email: terra@dft.if.uerj.br}} \address{ Departamento de
Astrof\'{\i}sica, Observat\'orio Nacional~--~CNPq, 
Rua General Jos\'e Cristino 77, S\~ao Crist\'ov\~ao, 20921-400 
Rio de Janeiro~--~RJ, Brazil}
\author{A.Z. Wang\thanks{Email:
wang@dft.if.uerj.br} }  
\address{ Departamento de F\' {\i}sica Te\' orica,
Universidade do Estado do Rio de Janeiro,
Rua S\~ ao Francisco Xavier 524, Maracan\~a,
20550-013 Rio de Janeiro~--~RJ, Brazil}

\date{\today}

\maketitle

\begin{abstract}

 The dynamics of collapsing and expanding cylindrically symmetric
gravitational and matter fields with lightlike wave-fronts  is studied 
in General Relativity, using the Barrab\'es-Israel method.  As an application
of the general formulae developed,   the collapse of a matter field that
satisfies the condition $R_{AB}g^{AB} = 0, \; (A, B = z, \varphi)$, in {\em an
otherwise flat spacetime background} is studied. In particular, it is found
that    the gravitational collapse of a purely gravitational  wave or a null
dust fluid cannot be realized in a  flat spacetime background.  The studies 
are further specified to the collapse of purely gravitational waves and
the general conditions for such collapse are found. It is shown that after the
waves arrive at the axis, in general, part of them is reflected to spacelike
infinity along the future light cone, and part of it is focused to form  
spacetime singularities on the symmetry axis. The cases where   the collapse
does not result in the formation of spacetime singularities are also
identified.  

\end{abstract}
 
\vspace{0.8cm}


\vskip2pc]
 
\section{Introduction}
 
 Gravitational collapse of a realistic body has been one of the most
thorny and important problems in Einstein's theory of
General Relativity.   Particularly, 
in 1991 Shapiro and Teukolsky \cite{ST91} studied  numerically the
problem of a dust spheroid, and found that only the spheroid is compact
enough, a black hole can be formed.  Otherwise, the collapse most
likely ends with a naked singularity in violation of Penrose's cosmic
censorship conjecture \cite{Penrose69}. Later,   by studying the collapse of a
cylindrical shell that is free of both gravitational and matter radiation and
is made of counter-rotating particles, Apostolatos and Thorne (AT) \cite{AT92}
showed analytically that the centrifugal forces associated with an arbitrarily
small amount of rotation, by themselves, without the aid of any pressure, can
halt the collapse at some non-zero, minimum radius, and the shell will then
oscillate until it settles down at some final, finite radius, whereby a
spacetime singularity is prevented from forming on the symmetry axis.  Soon
after AT's work, Shapiro and Teukolsky  \cite{ST92} studied numerically the
gravitational collapse of rotating spheroids, and found that the rotation
indeed significantly modifies the evolution when it is sufficiently large.
But, for small enough angular momentum, their simulations showed that
spindle singularities appeared to arise without apparent horizons, too. Hence,
it is still possible that even spheroids with some angular momentum naked
singularities may be formed. However, it should be noted that the
possibility of the formation of a horizon at late times cannot be excluded in
the ST simulations, since their numerical calculations always terminate right
at the formation of the spacetime singularities. 

Quite recently, we have studied analytically the problem  of a  collapsing
cylindrical shell that radiates gravitational waves and massless particles
when it is collapsing, and found two physically distinguishable final states
\cite{PW00}. In one case, after the shell collapses to a minimal non-zero
radius, it starts to expand, that is, the angular momentum of the dust
particles is strong enough to halt the collapse,  so that  a spacetime
singularity is prevented from forming on the symmetry axis.   However, in  the
other case the rotation is not strong enough to halt the collapse at a finite
non-zero radius, and, as a result, a spacetime singularity is finally formed on
the symmetry axis.  

As shown numerically in \cite{ST91,ST92}, during the late stages of the
collapse, the spheroid tends to fall at nearly the speed of light. Therefore,
it is reasonable to consider the collapsing body as consisting of null fluid
at its late stages of collapse. In this paper, we shall generalize our previous
studies \cite{PW00} to the   gravitational collapse of a cylinder
with a light-like wave-front.  Specifically, the paper is organized as
follows: In Sec. II, we consider the general matching of two
cylindrically symmetric regions across a null hypersurface, using the
Barrab\'es-Israel (BI) method \cite{BI91}, with the possibility that on the
null hypersurface a null dust shell may appear.  It should be noted that  the
same problem was also studied in \cite{LW94} but in double null
coordinates. As we shall see below, the method used there is much simpler 
than the BI one to be used  in this paper, but the advantage of the latter is
that it allows the independent choice of coordinates in both sides of the
hypersurface, and that the use of the radial coordinate in the BI method makes
 physical interpretations of the problem more transparent. In Sec. III
we apply the general formulae to the case where inside the null hypersurface
the spacetime is Minkowskian, while outside the shell the spacetime
satisfies the condition $R_{AB}g^{AB} = 0, \; (A, B = z, \varphi)$. In Sec.
IV, we apply the general formulae    further to the
collapse of cylindrically symmetric purely gravitational waves, while in Sec.
V our main conclusions are presented.

\section{Dynamics of Cylindrical Null Shells Without Rotation}

\renewcommand{\theequation}{2.\arabic{equation}}
\setcounter{equation}{0}

Both static \cite{stashell} and  dynamic \cite{PW00,dyshell} cylindrically
symmetric and timelike thin shells with zero total angular momentum have been
studied previously. In this section, we shall give a general treatment for
cylindrically symmetric null shells, which connects two cylindrical,
otherwise, arbitrary regions. In some cases the null shell can be made
disappear, and the corresponding null hypersurface becomes a boundary surface
\cite{Israel66}. In the following we shall consider the latter as a particular
case of the former and study them all together.

Assume that the shell, located on the hypersurface $\Sigma$,
divides the whole spacetime into two regions, $V^{\pm}$. Let  $V^{+}$ denote
the region outside the shell,  and $V^{-}$ denote the region inside the shell.
In $V^{+}$, the  metric takes the form   
\bqn
\label{2.1}  
ds^2_{+} &=& f^{+}(T, R){dT}^ 2 -
g^{+}(T, R){dR}^{2} - h^{+}(T, R)dz^{2}\nb\\
& & - l^{+}(T, R)d\varphi^{2},\; 
\left(R \ge R_{0}(T)\right),
\eqn
where $\{x^{+\mu}\} = \{T, \; R, \; z, \; \varphi\}$ with $- \infty < T, z < +
\infty, \; 0 \le \varphi \le 2\pi$,  and $R = R_{0}(T)$ is the location of the
shell in the coordinates $x^{+ \mu}$. If the sources   are
confined within a finite region in the radial direction, then the spacetime
should be asymptotically flat in the radial direction as ${\cal{R}}\rightarrow 
\infty$, where  ${\cal{R}}$ denotes the proper  radial distance. 

In $V^{-}$, the  metric takes the form   
\bqn 
\label{2.2} 
ds^2_{-} &=& f^{-}(t,
r){dt}^ 2 - g^{-}(t, r){dr}^{2} - h^{-}(t, r)dz^{2}\nb\\
& &  - l^{-}(t, r)d\varphi^{2},\; 
\left(r \le r_{0}(t)\right),
\eqn
where $\{x^{-\mu}\} = \{t, \; r, \; z, \; \varphi\}$ are the
usual cylindrical coordinates chosen in the region $V^{-}$. In this region we
shall choose the radial coordinate $r$ such that $r  = 0$ represents the
symmetry axis.  The hypersurface $\Sigma$ in this region is described by $r =
r_{0}(t)$. To have the cylindrical symmetry be well defined, the spacetime  in
this region has to satisfy several conditions \cite{PSW96,MS98,CSV99}:

(i) {\em The existence of an axially symmetric axis}: The spacetime that has
an axially symmetric axis is assured by the condition,
\bq
\lb{2.2a}
||\partial_{\varphi}|| = |g_{\varphi\varphi}| \rightarrow O(r^{2}),
\eq
as $r \rightarrow 0^{+}$.  

(ii) {\em The elementary flatness on the axis}: This condition requires that
the spacetime be locally flat on the axis, which in the present case can be
expressed as \cite{Kramer80}
\bq
\lb{2.2b}
\frac{X_{,\alpha}X_{,\beta}g^{\alpha\beta}}{4X}  \rightarrow 1,
\eq
as $r \rightarrow 0^{+}$, where $X$ is given by $X = ||\partial_{\varphi}|| =
|g_{\varphi\varphi}|$.    

Since the trajectory of the shell is null, we have 
\bqn
\lb{2.3a}
R'{}_{0}(T) &=& \left.\epsilon \left(\frac{f^{+}}
{g^{+}}\right)^{1/2}\right|_{R = R_{0}(T)}, \nb\\
r'{}_{0}(t) &=& \left.\epsilon \left(\frac{f^{-}}
{g^{-}}\right)^{1/2}\right|_{r = r_{0}(t)},
\eqn
where a prime denotes the ordinary differentiation with respect to the
indicated argument, and $\epsilon = \pm 1$, with the sign ``+" corresponding
to expanding shells, and the sign ``$-$" to collapsing shells.  On the shell,
the intrinsic coordinates will be chosen as  $\{\xi^{a}\} = \{w, \; z, \;
\varphi\},\; (a = 1,2,3)$, where $w$ denotes a null coordinate, which will be
defined explicitly below. In terms of $\xi^{a}$, the metric on the shell takes
the form  
\bq  
\label{2.3}
\left.ds^2\right|_{\Sigma}=  g_{* \; ab}d\xi^{a}d\xi^{b} =    -
h(w)dz^{2} - l(w)d\varphi^{2},
\eq
where $g_{* \; ab}$ denotes the induced metric on the hypersurface.
The first junction condition that the metrics in both sides of the shell
reduce to the same metric (\ref{2.3}) on $\Sigma$ requires that
\bqn
\label{2.4}
h(w) &\equiv& h^{+}(T, R_{0}(T)) = h^{-}(t, r_{0}(t)),\nb\\
l(w) &\equiv& l^{+}(T, R_{0}(T)) = l^{-}(t, r_{0}(t)),
\eqn
where on the hypersurface $\Sigma$ we have $T = T(w)$ and $t = t(w)$.
The normal vector to the hypersurface $\Sigma$ is given in $V^{+}$ and
$V^{-}$, respectively, by
\bqn
\lb{2.5}
n^{+}_{\mu} &=& \left\{- R'{}_{0}(T), \; 1, \; 0,\; 0\right\},\nb\\
n^{-}_{\mu} &=& \left\{- r'{}_{0}(t), \; 1, \; 0,\; 0\right\}.
\eqn
Following Barrab\'es and Israel (BI) \cite{BI91}, let us introduce the vectors
$N^{\pm}_{\mu}$ via the relations,
\bq
\lb{2.6}
N^{\pm \lambda} N^{\pm}_{\lambda} = 0, \;\;\;\;
N^{\pm \lambda} n^{\pm}_{\lambda} = 1,
\eq
we find that in the present case they take the form, 
\bqn
\lb{2.7}
N^{+ \mu} &=& \frac{1}{2}\left\{- \epsilon \left(\frac{g^{+}}
{f^{+}}\right)^{1/2} \delta^{\mu}_{T} + \delta^{\mu}_{R}\right\},\nb\\ 
N^{-\mu} &=& \frac{1}{2}\left\{- \epsilon \left(\frac{g^{-}}
{f^{-}}\right)^{1/2}\delta^{\mu}_{t} + \delta^{\mu}_{r}\right\}. 
\eqn
Defining the quantities, $e^{\pm\mu}_{(a)}$ by  $e^{\pm\mu}_{(a)} \equiv 
\partial x^{\pm \mu}/\partial \xi^{a}$, we obtain   
\bqn 
\lb{2.8}
e^{+\mu}_{(1)} &=& \frac{dT}{dw}\left\{\delta^{\mu}_{T} + R'{
}_{0}(T)\delta^{\mu}_{R}\right\},\nb\\
 e^{+\mu}_{(2)} &=& 
\delta^{\mu}_{z},\;\;\; e^{+\mu}_{(3)} =  \delta^{\mu}_{\varphi},\nb\\
e^{+\mu}_{(1)} &=&  \frac{dt}{dw}\left\{\delta^{\mu}_{t} +
r'{}_{0}(t)\delta^{\mu}_{r}\right\},\nb\\
e^{+\mu}_{(2)} &=&  \delta^{\mu}_{z},\;\;\;
e^{+\mu}_{(3)} =  \delta^{\mu}_{\varphi}.
\eqn
To make sure that $N^{\pm}_{\mu}$, defined on the two faces of the
hypersurface $\Sigma$, do represent the same vector $N^{\mu}$,  
we impose the conditions \cite{BI91}
\bq
\lb{2.9}
\left. N_{\lambda}^{+}e^{+ \lambda}_{(a)}\right|_{\Sigma^{+}} = 
\left. N_{\lambda}^{-}e^{- \lambda}_{(a)}\right|_{\Sigma^{-}}, 
\; (a = 1,\; 2,\; 3).
 \eq
In the present case, it can be shown that the above conditions reduce to
\bq
\lb{2.10}
\left.\left(f^{+}g^{+}\right)^{1/2} \frac{dT}{dw}\right|_{\Sigma^{+}} = 
\left.\left(f^{-} g^{-}\right)^{1/2} \frac{dt}{dw}\right|_{\Sigma^{-}}.
\eq
The above equation can be also considered as defining
the relation between $T$ and $t$ on the hypersurface.   
Following BI \cite{BI91}, we define the extrinsic curvature of the
hypersurface $\Sigma$ as,
\bqn
\lb{2.11}
{\cal{R}}_{ab} &\equiv& - N_{\lambda}e^{\nu}_{(b)}\left(\nabla_{\nu}
e^{\lambda}_{(a)}\right) \nb\\
&=& -N_{\lambda}e^{\nu}_{(b)}
\left\{e^{\lambda}_{(a),\nu} + \Gamma^{\lambda}_{\nu\mu}e^{\mu}_{(a)}\right\}, 
\eqn    
which in the present case has the following non-vanishing components,
\bqn
\lb{2.12}
{\cal{R}}^{+}_{ww} &=& \epsilon \left(f^{+}g^{+}\right)^{1/2}\left\{
\frac{d^{2}T}{dw^{2}} +
\frac{1}{2}\left(\frac{dT}{dw}\right)^{2}\left(\frac{f^{+}_{,T}}{f^{+}}
+ \frac{g^{+}_{,T}}{g^{+}}\right)\right\}\nb\\
& &  +
\left(\frac{dT}{dw}\right)^{2}f^{+}_{,R},\nb\\
{\cal{R}}^{+}_{zz} &=&
\frac{1}{4}\left\{\epsilon\left(\frac{f^{+}}{g^{+}}\right)^{1/2}  h^{+}_{,T} -
h^{+}_{,R}\right\},\nb\\ 
{\cal{R}}^{+}_{\varphi\varphi} &=&
\frac{1}{4}\left\{\epsilon\left(\frac{f^{+}}{g^{+}}\right)^{1/2}  l^{+}_{,T} -
l^{+}_{,R}\right\},
\eqn
where $f^{+}_{,T} \equiv \partial f^{+}(T,R)/\partial T$ etc., and
${\cal{R}}^{-}_{ab}$ can be obtained from the above expressions by the replacement
\bqn
\lb{eq2.13}
f^{+},& & g^{+},\; h^{+},\; l^{+},\; R_{0}(T),\; T,\; R \nb\\
& &  \rightarrow
\;\; f^{-},\; g^{-},\; h^{-},\; l^{-},\; r_{0}(t),\; t,\; r. 
\eqn
Introducing the null vector $l^{a}$ along the direction $dw$ by $l^{a} =
\delta^{a}_{w}$, it can be shown that the surface energy-momentum tensor
$\tau_{ab}$ is given by \cite{BI91},
\bqn
\lb{2.14}
\tau^{ab} &=& - \frac{1}{2\kappa}\left\{g^{ac}_{*}l^{b}l^{d} +
g^{bd}_{*}l^{a}l^{c} \right.\nb\\
& & \left. - g^{ab}_{*}l^{c}l^{d} -
g^{cd}_{*}l^{a}l^{b}\right\}\gamma_{cd}, 
\eqn
where $\kappa \equiv 8\pi G/c^{4}$,
\bq
\lb{2.15}
\gamma_{ab} \equiv 2\left({\cal{R}}^{+}_{ab} - {\cal{R}}^{-}_{ab}\right),
\eq
and
\bqn
\lb{2.15a}
\left[g_{*\; ab}\right] &=& \left[\matrix{0& 0 & 0\cr
0 & - h(w) & 0 \cr
0 & 0& -l(w)\cr}\right],\nb\\
\left[g^{ab}_{*}\right]&=& \left[\matrix{0& 0 & 0\cr
0 & - h^{-1}(w) & 0 \cr
0 & 0& - l^{-1}(w)\cr}\right].
\eqn
Inserting Eqs.(\ref{2.10}), (\ref{2.12}) and the corresponding expressions for
${\cal{R}}^{-}_{ab}$ into the above equations, we find that $\tau^{ab}$ can be
written in the form
\bq
\lb{2.16}
\tau^{ab} = \rho l^{a}l^{b} + p \left(z^{a}z^{b} 
+ \varphi^{a}\varphi^{b}\right), \; (a, b = w, \; z, \; \varphi),
\eq
where 
\bqn
\lb{2.17}
\rho &=& - \frac{1}{2\kappa}\left[ 
\frac{\gamma_{zz}}{h(w)} + \frac{\gamma_{\varphi\varphi}}{l(w)}\right]\nb\\
&=& -
\frac{1}{4\kappa}\left\{\epsilon\left[\left(\frac{g^{+}}{f^{+}}\right)^{1/2}
\left(\frac{h^{+}_{,T}}{h^{+}} +
\frac{l^{+}_{,T}}{l^{+}}\right)\right.\right.\nb\\ 
& & \left. -  
\left(\frac{g^{-}}{f^{-}}\right)^{1/2} \left(\frac{h^{-}_{,t}}{h^{-}} +
\frac{l^{-}_{,t}}{l^{-}}\right)\right] \nb\\  
& & -
\left. \left[\left(\frac{h^{+}_{,R}}{h^{+}} +
\frac{l^{+}_{,R}}{l^{+}}\right) - \left(\frac{h^{-}_{,r}}{h^{-}} +
\frac{l^{-}_{,r}}{l^{-}}\right)\right]\right\},\nb\\  
p &=& -
\frac{\gamma_{ww}}{2\kappa}\nb\\ &=&-
\frac{1}{2\kappa}\left(\frac{dT}{dw}\right)^{2}\left\{
\frac{1}{g^{+}}\left(g^{+}f^{+}_{,R} - f^{+}g^{+}_{,R}\right)\right.\nb\\
& & \left. -
\frac{f^{+}g^{+}}{g^{-}}\left(\frac{f^{-}_{,r}}{f^{-}}  -
\frac{g^{-}_{,r}}{g^{-}}\right)\right\},   
\eqn
and 
\bq
\lb{2.18}
z_{a} = h^{1/2}(w)\delta^{z}_{a},\;\;\;\;\;
\varphi_{a} = h^{1/2}(w)\delta^{\varphi}_{a}.
\eq
Clearly, the surface energy-momentum tensor given by Eq.(\ref{2.16})
represents a Type II  null fluid in the classifications of Hawking and
Ellis \cite{HE73}, where the null fluid is moving along the lines defined by
the tangential vector $l^{a}$ with its isotropic pressure $p$ in the $dz$
and $d\varphi$ directions. It is interesting to note that when $f^{+} = g^{+}$
and  $f^{-} = g^{-}$, we have $p = 0$.

\section{Gravitational Collapse of Matter Fields Satisfying 
     $R_{AB}g^{AB} = 0$}

\renewcommand{\theequation}{3.\arabic{equation}}
\setcounter{equation}{0}

As an application of the formulae developed above, let us consider some
specific models that represent gravitational collapse. Therefore, from now on
we shall consider only the case where $\epsilon = -1$. The wave front of a
collapsing cylinder is given by $R = R_{0}(T)$ in the $x^{+\mu}$ coordinates
and $r = r_{0}(t)$ in the $x^{- \mu}$ coordinates. Before the wave arrives, we
assume that the spacetime is flat, that is,  
\bq 
\lb{3.1}
ds^{2}_{-} = dt^{2} - dr^{2} - dz^{2} - r^{2}d\varphi^{2}, \; 
\left(r \le r_{0}(t)\right).
\eq
Clearly, in the present case the cylindrical symmetry of the spacetime is well
defined, and the axis $r = 0$ is free of any kind of spacetime singularities
[cf. Fig. 1(a)]. 

\begin{figure}[htbp]
  \begin{center}
    \leavevmode
    \epsfig{file=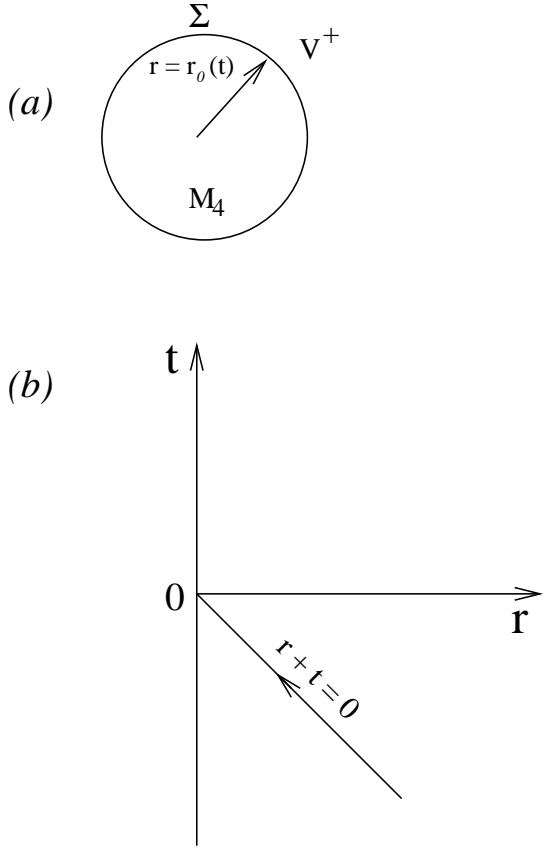,width=0.4\textwidth,angle=0}
\caption{The matching of a Minkowskian region $M_{4}$ with an exterior
$V^{+}$: (a) In the $z = Const.$ plane. The hypersurface
$\Sigma$ is a null surface represented by the circle $ r = r_{0}(t)$. (b) In
the ($t, \; r$) -plane. The hypersurface $\Sigma$ now is represented by the
straight line $ t + r = 0$. In the region $t + r < 0$, the spacetime is
Minkowskian, while in the region $t + r > 0$, the spacetime is curved.}     
 \label{fig1}   
\end{center} 
\end{figure}


Then,  the condition (\ref{2.3a}) yields,  
\bq 
\lb{3.2}
r_{0}(t) = - (t - t_{0}),
\eq
where $t_{0}$ is a constant. By a translation transformation in $t$, we
can always set  $t_{0} = 0$. In the following,   we shall always assume that
this is the case. Since $r_{0}(t) \ge 0$, we must have  $t \le 0$
on the hypersurface $\Sigma$. Then, we can see that the history of this hypersurface
 in the ($t, r$)-plane is represented by the part of the  light cone
$t + r = 0, \; t \le 0$, as shown by Fig. 1(b).

In the region $R \ge R_{0}(T)$, where collapsing
cylinder is present, the metric, without loss of generality, can be cost in
the form   
\bqn
\lb{3.3}
ds^{2}_{+} &=& e^{-M}\left(dT^{2} - dR^{2}\right) \nb\\
& & - e^{-U}\left(e^{V}dz^{2} + e^{-V}d\varphi^{2}\right),
\eqn
where $M, \; U$ and $V$ are functions of $T$ and $R$ only. Then, 
Eqs.(\ref{2.3a}), (\ref{2.4}) and (\ref{2.10}) yield
\bqn
\lb{3.3a}
R_{0}(T) &=& - T  \ge 0,\;\;
V\left(T, R_{0}(T)\right) = U\left(T, R_{0}(T)\right),\nb\\
r_{0}(t) &=& e^{-U\left(T, R_{0}(T)\right)},\;\;
dt = e^{-M\left(T, R_{0}(T)\right)} dT.
\eqn
Inserting the above equations into Eq.(\ref{2.17}), we find that 
\bq
\lb{3.3b}
\rho =  - \frac{1}{2\kappa}\left.\left\{ U_{,T} + U_{,R} +
e^{U}\right\}\right|_{R = - T},\;\;\;\;
p = 0.
\eq
 To further study the problems, in the following we shall consider
only the case where the cylinder is made of matter that satisfies the
condition 
\bq 
\lb{3.4} 
R_{AB} g^{AB} = 0,\; (A, B = z, \; \varphi),
\eq
where $R_{\mu\nu}$ denotes the Ricci tensor. It can be shown that several
matter fields, such as, null dust fluid, massless scalar field, and
electromagnetic field, satisfy the above condition \cite{Kramer80,WIM79}. For
the metric given by (\ref{3.3}), Eq.(\ref{3.4}) yields
\bq
\lb{3.5}
U_{,TT} - U_{,T}^{2} - U_{,RR} + U_{,R}^{2} = 0,
\eq
which has the general solution
\bq
\lb{3.6}
U = - \ln\left[a(T+R) + b(T-R)\right],
\eq
where $a$ and $b$ are arbitrary functions of their indicated
arguments, subject to 
\bq
\lb{3.7}
a + b \ge 0.
\eq
Inserting the above expressions into Eq.(\ref{3.3b}), we find that
\bq
\lb{3.8}
\rho = \left.\frac{2a'  - 1}{2\kappa(a + b)}\right|_{R = - T}, \;\;\;\;
p = 0.
\eq
According to 
\bqn
\lb{3.9}
&a)& \; a'b' <  0,\;\;\;\; 
b) \; a'b' > 0,\nb\\
&c)& \; a' = 0, \;\;  b' \not= 0,\nb\\
&d)& \; a' \not= 0, \;   b' = 0,
\eqn
the solutions will have physically different interpretations, thus, in the
following let us consider them separately.

{\bf Case a)} $\; a'b' < 0$: In this case the normal vector to the
hypersurfaces $\Phi \equiv e^{-U} =  Const.$ is spacelike. As a matter of
fact, it is easy to show that
\bq
\lb{3.10}
\Phi_{,\alpha}\Phi_{,\beta}g^{\alpha\beta} = 4 a'b'e^{M} < 0.
\eq
Without loss of generality, now we can  choose 
\bq
\lb{3.13}
a = \frac{1}{2}(T + R),\;\;\;\;
b = -\frac{1}{2}(T - R).
\eq
In fact, introducing two new coordinates  $\bar{T}$ and $\bar{R}$ via the
relations,  
\bq
\lb{3.11}
\bar{R} = a + b,\;\;\;\; \bar{T} = a - b,
\eq
we find that the corresponding metric takes the form,
\bq
\lb{3.12}
ds^{2}_{+} = e^{-\bar{M}}\left(d\bar{T}^{2} - d\bar{R}^{2}\right) -
\bar{R}\left(e^{V} dz^{2} + e^{-V}d\varphi^{2}\right),
\eq
where $e^{-\bar{M}} \equiv e^{-M}/(-4a'b')$, and $\bar{M}$ and $V$ now are
functions of $\bar{T}$ and $\bar{R}$ only. Clearly, the above form of metric
is the same as that given by Eqs.(\ref{3.3}) with the choice of  
Eq.(\ref{3.13}), from which we find that
\bq
\lb{3.12a}
e^{- U} = a + b = R.
\eq
Combining Eqs.(\ref{3.2}) and (\ref{3.3a}), we find that the first junction
condition now requires
\bqn
\lb{3.12b}
r_{0}(t) &=& R_{0}(T)= -t = -T \ge 0,\nb\\
V\left(T, R_{0}(T)\right) &=& - \ln \left[R_{0}(T)\right],\;\;
M\left(T, R_{0}(T)\right) = 0.
\eqn
Substituting Eq.(\ref{3.13}) into Eq.(\ref{3.8}), on the other
hand, we find that 
\bq
\lb{3.13a}
\rho = 0 = p. 
\eq
That is, in this case the matching across the hypersurface $\Sigma$ is
always smooth, and no matter shell appears on the hypersurface. 
Using the ${\cal{C}}$-energy defined by Thorne \cite{Thorne65}, 
\bq
\lb{3.12c}
{\cal{C}}_{E} \equiv \frac{1}{8 }\left(1 - \frac{e^{M-V}}{R}\right),
\eq
we find from Eq.(\ref{3.12b}) that on the hypersurface $\Sigma$ we have 
\bq
\lb{3.12d}
{\cal{C}}_{E}\left(T, R_{0}(T)\right) = 0.
\eq
That is, the total ${\cal{C}}$-energy inside the cylinder $R = R_{0}(T)$  is
zero. This is what one would expect, as the hypersurface $\Sigma$ now is free
of matter shells and the spacetime, before the wave front arrives, is flat.

{\bf Case b)} $\; a'b' > 0$: In this case it can be shown that the normal
vector to the hypersurfaces $\Phi \equiv e^{-U} =  Const.$  is timelike.
Following a similar argument as that given in the last case, one can show that
now we can  choose  
\bq
\lb{3.14}
a = -\frac{1}{2}(T + R),\;\;\;\;
b =  -\frac{1}{2}(T - R),
\eq
so that  
\bq
\lb{3.15}
e^{-U} = a + b = - T \ge 0.  
\eq
Inserting Eq.(\ref{3.14}) into Eq.(\ref{3.8}), we find  
\bq
\lb{3.16}
\rho = - \frac{1}{\kappa (-T)},\;\;\; p = 0.
\eq
Since on the hypersurface $\Sigma$ we have $T \le 0$, from the above we can see
that  the energy density $\rho$ is always negative. This  is not physical and
we shall  not consider this case further in the following discussions.    

{\bf Case c)} $\; a' = 0,\; b' \not= 0$: In this case, without loss of
generality, we can set $a = 0$. Then, it can be shown  that the hypersurfaces
$\Phi \equiv e^{-U} = b(T - R) = Const.$ now become null, and the function $ b$ is 
positive and otherwise arbitrary.   Hence, from
Eq.(\ref{3.8}) we find   
\bq
\lb{3.17}
\rho = - \frac{1}{2\kappa b(2R)}, \;\;\; p = 0,
\eq
from which we can see that the energy density $\rho$ is always negative,
too, and this case should be also discarded in the
following discussions.

{\bf Case d)} $\; a' \not = 0,\; b' = 0$: Similar to the last case,
it can be shown that now the hypersurfaces $\Phi \equiv e^{-U} =  Const.$ are
also null, and the function $ a$ is  arbitrary, subject to $a \ge
0$, with the choice $b = 0$.  Then, the corresponding surface energy density 
is given by 
\bq 
\lb{3.18}
\rho = \frac{2a'(0) -1}{2\kappa a(0)}.
\eq
On the other hand, from Eq.(\ref{3.3a}) we find that
\bq
\lb{3.19}
r_{0}(t) = e^{-U\left(T, R_{0}(T)\right)} = a(0),
\eq
which is not consistent with Eq.(\ref{3.2}). Thus, this case has to be
discarded as representing gravitational collapse, too. 

It is interesting to note that in the last two cases the only
possible matter field is a null dust fluid with the energy-momentum tensor
(EMT) being given by
\bq
\lb{3.19a}
T_{\mu\nu} = \rho k_{\mu}k_{\nu},
\eq
where $k_{\mu}$ is a null vector. In the last case it is given by $k_{\mu} =
\delta^{T}_{\mu} + \delta^{R}_{\mu}$, and the corresponding EMT represents an
ingoing null dust fluid moving along the hypersurfaces $T + R = Const.$, while
in Case c), it is given by $k_{\mu} = \delta^{T}_{\mu} - \delta^{R}_{\mu}$, and
the corresponding EMT represents an outgoing  null dust fluid moving along the
hypersurfaces $T - R = Const.$ In both of the two cases, an, ingoing in the
last case, and outgoing in Case c),  purely  gravitational cylindrical  wave
is present, except for the case where $V = Const.$   

In review of all the above, we find that  {\em the gravitational
collapse in the last three cases in an otherwise flat and cylindrical spacetime
background   cannot be  realized}.  The only case that is relevant to
gravitational collapse  is Case a). This result is a little bit of surprise,
and mainly due to the requirement that, before the wave front arrives, the
spacetime be flat and take the cylindrical form  Eq.(\ref{3.1}). Although 
Birkhoff's theorem for the spherically symmetric case is not applicable to the
cylindrically symmetric case \cite{Kramer80},  from the physical point of view
this requirement seems quite reasonable.  

To further study the problem, let us now turn to consider gravitational
collapse of cylindrical purely gravitational  waves.

\section{Gravitational Collapse of Purely Gravitational Waves}

\renewcommand{\theequation}{4.\arabic{equation}}
\setcounter{equation}{0}

As shown in the last section,  the only case that is relevant to
gravitational collapse in an otherwise flat background is Case a), in which
the null hypersurface $r =r_{0}(t)\;$ [ or $\; R =R_{0}(T)$] is free of  
matter, $\tau_{ab} = 0$, and the function $U(T, R)$ is chosen as 
\bq 
\lb{3.20} 
U = - \ln R.
\eq
For this gauge, it can be shown that only three of the vacuum Einstein field
equations $R_{\mu\nu}= 0$ are independent and can be written in the form
\bqn
\lb{3.23a}
\Omega_{,R} &=& \frac{R}{2}\left(V_{,T}^{2} + V_{,R}^{2}\right),\\
\lb{3.23b}
\Omega_{,T} &=& R V_{,T}  V_{,R},  \\
\lb{3.23c}
V_{,RR} &+&  \frac{1}{R} V_{,R} - V_{,TT} =0,
\eqn
where 
\bq
\lb{3.22}
\Omega(T, R) = \frac{1}{2}\ln R - M(T, R).
\eq
Thus, once the solution of Eq.(\ref{3.23c}) is known, the function
$\Omega$ can be obtained by the quadrature from Eqs.(\ref{3.23a}) and
(\ref{3.23b}). The general solution of Eq.(\ref{3.23c}) is given by
\cite{Wang93,Verdaguer93} 
\bqn
\lb{3.25}
V &=&  \int^{\infty}_{0}{\left[A(\omega)\sin(\omega T) +
B(\omega)\cos(\omega T)\right]J_{0}\left(\omega R\right)}
d\omega\nb\\
& +&  \int^{\infty}_{0}{\left[C(\omega)\sin(\omega T) +
D(\omega)\cos(\omega T)\right]N_{0}\left(\omega R\right)}
d\omega\nb\\ 
&+ & \alpha T + \beta \ln R + \sum^{N}_{k =
1}{h_{k}\ln\left(\frac{\mu_{k}}{R}\right)},    
\eqn
where 
\bq
\lb{3.26}
\mu_{k} \equiv T_{k} - T \pm \left[\left(T_{k} - T\right)^{2} -
R^{2}\right]^{1/2},
\eq
$\alpha,\; \beta,\; A(\omega),\; B(\omega),\; C(\omega),\; \;D(\omega),\;
\omega,\; h_{k}$ and $T_{k}$ are arbitrary constants, and $N$ is an integer.
The functions  $J_{0}\left(\omega R\right)$ and
$N_{0}\left(\omega R\right)$ denote the Bessel and Neumann functions of 
zero order, respectively. The integration constants $T_{i}$ can be real or
complex, and in general takes the form 
 \bq
\lb{3.27}
T_{k} = T^{0}_{k} + i \; w_{k},\; (k = 1,\; 2,\; ...,\; N), 
\eq
where $T^{0}_{k}$'s and $w_{k}$'s are real. If one starts with a complex term,
say, $\mu_{j}$, then its complex conjugate $\mu^{*}_{j}$  should be also
included, in order to make the metric coefficients to be real. Note that when
$T_{k}$ is real, the solution is defined only in the regions $(T_{k} - T)^{2}
\ge R^{2}$ [cf. Fig. 2].


\begin{figure}[htbp]
  \begin{center}
    \leavevmode
    \epsfig{file=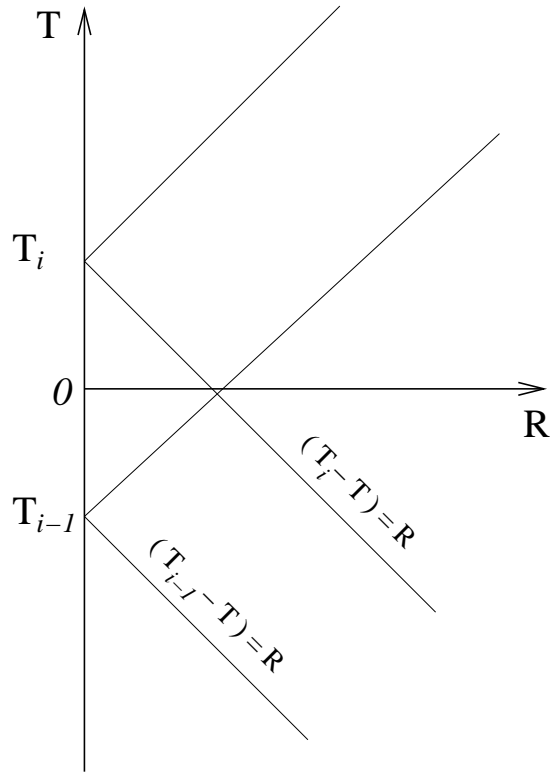,width=0.4\textwidth,angle=0}
\caption{The function $\mu_{k}$ appearing in the expression (\ref{3.25}) is
defined only in the regions $(T_{k} - T)^{2} \ge R^{2}$ when
$T_{k}$ is real. On the light cone, we have $\mu_{k} = R$.}   
\label{fig2}  
\end{center}
\end{figure}

 
On the light cone $(T_{k} - T)^{2}
= R^{2}$, we have $\ln(\mu_{k}/R) = 0$. Then, from Eq.(\ref{3.25}) we can see
that $V(T,R)$ matches continuously to the solution that does not include
$\mu_{k}$ in the region $(T_{k} - T)^{2}
\le R^{2}$.  Belinskii
and Francaviglia interpreted this as gravitational shock-wave solutions
\cite{BF82}, but later Gleiser  showed that in general the
light cone also support a null dust fluid  \cite{Gleiser84}. Yet,  Curir {\em
et al.} found that sometimes the solutions have curvature singularities on the
light cone \cite{CFV92}.   In any case,
since here we are mainly concerned with gravitational collapse of
gravitational waves only, we shall not consider this possibility any more, and
simply assume that all the $\mu_{k}$'s are  complex, as in the latter case
there are no such restrictions and the solutions are valid in the whole
spacetime. The term related to $\mu_{k}$ is usually referred to as generalized
soliton solutions \cite{GV87}.

On the hypersurface $\Sigma$ where $T = - R$, the first junction condition 
Eq.(\ref{3.12b}) requires
\bq
\lb{3.27a}
V\left(-R, R\right) = - \ln R,\;\;\; M\left(-R, R\right) = 0.
\eq
Since for the general solution of Eq.(\ref{3.25}), the function $M(T, R)$ is
not known, in the following let us consider the above condition only in the two
limits: $R \rightarrow 0$ and $R \rightarrow \infty$. When $R \rightarrow 0$,
we have
\bqn
\lb{3.28}
J_{0}(\omega R) &\rightarrow& 1,\nb\\
N_{0}(\omega R) &\rightarrow& \frac{2}{\pi}\ln\left(\frac{\omega
R}{2}\right),\nb\\
\mu^{+}_{k}  &\rightarrow& 2T_{k},\nb\\
\mu^{-}_{k}  &\rightarrow& \frac{R^{2}}{2T_{k}}.
\eqn
We assume, without loss of generality, that in  
Eq.(\ref{3.25}) the first $n$ $\mu_{k}$'s take the values of $\mu^{-}_{k}$ and
the rest $(N-n)$ $\mu_{k}$'s take the values of $\mu^{+}_{k}$.  
Then we find that
\bq
\lb{3.30}
V(-R, R) \rightarrow (\beta + h^{-} + d) \ln R,
\eq
as $R \rightarrow 0$, where
\bq
\lb{3.31}
h^{-} \equiv \sum^{n}_{k = 1}{h^{-}_{k}},\;\;\;
d \equiv \frac{2}{\pi}\int^{\infty}_{0}{D(\omega)d\omega}.
\eq
The junction condition Eq.(\ref{3.30})  requires   
\bq
\lb{3.32}
\beta + h^{-} + d  = -1.
\eq
When $R \rightarrow \infty$, we have
\bqn
\lb{3.28a}
J_{0}(\omega R) &\rightarrow& \left(\frac{2}{\pi \omega R}\right)^{1/2}
\cos\left(\omega R - \frac{\pi}{4}\right),\nb\\\
N_{0}(\omega R) &\rightarrow& \left(\frac{2}{\pi \omega R}\right)^{1/2}
\sin\left(\omega R - \frac{\pi}{4}\right),\nb\\
\mu^{\pm}_{k}  &\rightarrow& R,
\eqn
 from which we find that
\bq
\lb{3.33}
V(-R, R) \rightarrow - \alpha R + \beta \ln R.
\eq
Thus, the first junction condition Eq.(\ref{3.27a}) 
in this limit requires that   
\bq 
\lb{3.34}
\alpha = 0, \;\;\;\; \beta = -1.
\eq
Combining  Eqs.(\ref{3.32}) and (\ref{3.34}) we find that 
\bq
\lb{3.35}
\alpha  = 0, \;\;\;\;
\beta = -1, \;\;\; d + h^{-} = 0.
\eq

To further study the problem, now let us turn to study the spacetime outside 
the hypersurface $\Sigma$, i.e., region $V^{+}$. In particular, following
\cite{GV87}, we shall concentrate ourselves in the four different regions:  
(i) The spacelike infinity, $|T| \ll R \rightarrow \infty$; (ii) the future
null infinity, $R \sim T \rightarrow \infty$;  (iii) the timelike future
infinity, $R \ll  T \rightarrow \infty$; and (iv) the focusing region, $R
\sim T \rightarrow 0^{+}$ [cf. Fig. 3].   

(i) {\em The spacelike infinity, $|T| \ll R \rightarrow \infty$}: In this
limit, let us first consider the first two terms in Eq.(\ref{3.25}), which
will be denoted by $V_{FB}(T, R)$. Using Eq.(\ref{3.28a}), we find that
\cite{Adams82}
\bqn
\lb{3.36}
V_{FB} &\rightarrow& \left(\frac{1}{2\pi
R}\right)^{1/2}\int^{\infty}_{0}{\left\{E^{+}(\omega)\cos\left[\omega(R-T) +
\phi^{+}(\omega)\right]\right.}\nb\\
& & {\left. + E^{-}(\omega)\cos\left[\omega(R+T) +
\phi^{-}(\omega)\right]\right\}\frac{d\omega}{\omega^{1/2}}},
\eqn
where
\bqn
\lb{3.37}
E^{\pm}(\omega) &=& \left\{[A(\omega) \pm D(\omega)]^{2}+ [C(\omega) \mp
B(\omega)]^{2}\right\}^{1/2},\nb\\
\phi^{\pm}(\omega) &=& \tan^{-1}\left[\frac{C(\omega) \mp B(\omega)}
{A(\omega) \pm D(\omega)}\right].
\eqn
It is clear that, as $R \rightarrow \infty$, these two terms represent a
superposition of two traveling gravitational waves along the radial direction,
one is ingoing represented by the term $E^{-}(\omega)$ and the other is
outgoing represented by the term $E^{+}(\omega)$. When the waves have only a
single wavelength, say, $\lambda_{0}$, it can be shown that the corresponding
function $M(T, R)$ takes the form \cite{Adams82},
\bqn
\lb{3.36a}
M_{FB} &\rightarrow& \frac{1}{2}\ln R -
\frac{2\omega_{0}}{\pi}\left[\left(A^{2}_{0} + B^{2}_{0} + C^{2}_{0} +
D^{2}_{0}\right)R\right.\nb\\
&+& \left.  2(B_{0}C_{0} - A_{0}D_{0})T\right]\nb\\
&-& \frac{1}{2\pi}\left\{E^{-}_{0}\cos\left[2\omega_{0}(T + R) -
\theta^{-}_{0}\right] \right.\nb\\
& +& \left. E^{+}_{0} \cos\left[2\omega_{0}(R - T) -
\theta^{+}_{0}\right]\right\}, \eqn
where $\omega_{0} = 2\pi/\lambda_{0}$, 
\bq
\lb{3.36b}
\theta^{\pm}_{0} \equiv \tan^{-1}\left[\frac{2(A_{0}\pm D_{0})(C_{0}\mp B_{0})}
{(A_{0}\pm D_{0})^{2} - (C_{0} \mp B_{0})^{2}}\right].
\eq
and  $A_{0} \equiv A(\omega_{0})$, etc. It can be shown that the spacetime is
singular in this limit. Since we are considering gravitational collapse, the
more interesting case would be that the spacetime is asymptotically
flat   in the radial direction, and it  is possibly
singular only at $R = 0$ for $T > 0$. Then,   we are sure that such
singularities are indeed formed by the collapse.
Therefore, in the following we shall assume that these two terms vanish
identically, 
\bq
\lb{3.36c}
A(\omega) = B(\omega) = C(\omega) =  D(\omega) = 0.
\eq
Hence, from Eq.(\ref{3.31}) we find that $d = 0$. Substituting it into 
Eq.(\ref{3.35}) we obtain
\bq
\lb{3.36d}
h^{-} = \sum^{n}_{k = 1}h^{-}_{k} = 0.
\eq

 The third term in Eq.(\ref{3.25}) vanishes too,
because of the first junction condition (\ref{3.35}). Thus, in the following
let us concentrate ourselves only on the fourth and last (soliton) terms with
$\beta = -1$. When the soliton term vanishes, the spacetime becomes Minkowskian. 

As we mentioned previously, in this paper we consider only the case where
$T_{k}$ is complex. For the complex $T_{k}$ it is found convenient to
introduce the following quantities \cite{GV87},
\bq
\lb{3.38}
\mu^{+}_{k} = R \sigma^{1/2}_{k}e^{i \gamma_{k}},\;\;
\mu^{-}_{k}  =  \frac{R^{2}}{\mu^{+}_{k}} = R \sigma^{-1/2}_{k}e^{- i
\gamma_{k}}, 
\eq
where
\bqn
\lb{3.38a}
\sigma_{k} &=& L_{k} + (L_{k}^{2} - 1)^{1/2},\nb\\
L_{k} &=& \frac{L^{0}_{+}}{R^{2}} + \left[1 - \frac{2L^{0}_{-}}{R^{2}} +
\frac{\left(L^{0}_{+}\right)^{2}}{R^{4}}\right]^{1/2},\nb\\ 
L^{0}_{\pm} &\equiv& T_{k}^{2} \pm w^{2}_{k},\;\;\; T_{k} \equiv T^{0}_{k} -
T,\nb\\  \gamma_{k}&=& \cos^{-1}\left[\frac{2\sigma^{1/2}_{k}T_{k}}{R(1 +
\sigma_{k})}\right].  
\eqn
If we further assume that
\bqn
\lb{3.38b}
h^{-}_{k + n/2} &=& \left(h^{-}_{k}\right)^{*} = \alpha^{-}_{k} - i \;
\beta^{-}_{k},\nb\\
h^{+}_{k + (N+n)/2} &=& \left(h^{+}_{k + n}\right)^{*} =  \alpha^{+}_{k} - i
\; \beta^{+}_{k},
\eqn
where $\alpha^{\pm}$'s and $\beta^{\pm}$'s are real constants, we find that
\bqn
\lb{3.39}
V_{soliton} &=& - \ln R + \sum^{N}_{k =
1}{h_{k}\ln\left(\frac{\mu_{k}}{R}\right)}\nb\\ 
& =& - \ln R + \sum^{s}_{k =
1}{\left(\frac{g_{k}}{2}\ln \sigma_{k} +
f_{k}\gamma_{k}\right)},
\eqn
where $s \equiv N/2$, and
\bqn
\lb{3.39a}
g_{k} &\equiv& \cases{-2\alpha^{-}_{k}, & $ 1 \le k \le n/2$\cr
2\alpha^{+}_{k + n/2}, & $ n/2 +1 \le k \le N/2$\cr},\nb\\
f_{k} &\equiv& \cases{2\beta^{-}_{k}, & $ 1 \le k \le n/2$\cr
2\beta^{+}_{k+n/2}, & $ n/2 +1 \le k \le N/2$\cr}.
\eqn
In terms of $\alpha^{-}_{k}$'s, Eq.(\ref{3.36d}) takes the form
\bq
\lb{3.39aa}
\sum^{n/2}_{k = 1}{\alpha^{-}_{k}} = 0.
\eq

The general  solution (\ref{3.39}) was  studied in details in \cite{GV87},
thus in the following we shall only summarize its main properties, and for the
details, we refer readers to \cite{GV87}. In particular, it can be shown that
\bqn
\lb{3.39b}
\sigma_{k} &\rightarrow& 1 + \frac{2w_{k}}{R} + O(R^{-2}),\nb\\
\gamma_{k} &\rightarrow& \frac{\pi}{2} - \frac{T_{k}}{R} + O(R^{-3}),
\eqn
as $R\rightarrow \infty$. Thus, from the above we can see that the terms that are
proportional to $\sigma_{k}$ are negligible, while the terms that 
proportional to $\gamma_{k}$ have the following distributions to the metric
coefficients,
\bqn
\lb{3.40}
 V_{soliton} &\rightarrow& - \ln R + \sum^{s}_{k = 1}{f_{k}\gamma_{k}} ,\nb\\
 M_{soliton} &\rightarrow& - \frac{f^{2}}{2}\ln R,
\eqn
where
\bq
\lb{3.40a}
f\equiv \sum^{s}_{k = 1}{f_{k}} = 2\left(\sum^{n/2}_{k = 1}{\beta^{-}_{k}} +
\sum^{(N+n)/2}_{k = (n+2)/2}{\beta^{+}_{k}}\right).
\eq
Thus, in this limit the solution is singular unless $f = 0$.  With the same
reason as that given following Eq.(\ref{3.36b}), in the following we shall
assume that 
\bq
\lb{3.40b}
\sum^{n/2}_{k = 1}{\beta^{-}_{k}} +
\sum^{(N+n)/2}_{k = (n+2)/2}{\beta^{+}_{k}} = 0, \; ( f = 0).
\eq
Then, we can see that the solution will represent the gravitational collapse
of purely gravitational cylindrical waves with lightlike wave-front, on which
no dust shell appears. The  spacetime is asymptotically flat in
the radial direction and locally flat near the symmetry axis.

(ii) {\em The future null
infinity, $R \sim T \rightarrow \infty$}: In this limit, it can be shown that
\bqn 
\lb{3.41}
\sigma_{k} &\rightarrow& 1 + O\left(R^{-1/2}\right),\nb\\
\gamma_{k} &\rightarrow& \pi + O\left(R^{-1/2}\right),
\eqn
and that the metric coefficients behave like
\bqn
\lb{3.41a}
V(T, R)  &\rightarrow&  - \ln R  + O\left(R^{-1/2}\right),\nb\\
M(T, R) &\rightarrow&   O\left(R^{-1/2}\right).
\eqn
Thus, in this region the spacetime is
asymptotically Minkowskian, too. A detailed study of the Riemann tensor 
showed that the spacetime in this region is Petrov type N and represents an
outgoing gravitational wave along the light cone \cite{GV87}.

(iii) {\em The timelike future infinity, $R \ll  T \rightarrow \infty$}: The
study of the solutions in this region will give us some information about the
final state of the collapse. It can be shown that 
\bqn
\lb{3.42}
\sigma_{k} &\rightarrow& \frac{4T^{2}}{R^{2}},\nb\\
\gamma_{k} &\rightarrow& \pi + O\left(\frac{R^{2}}{4T^{2}}\right),
\eqn
and 
\bqn
\lb{3.42a}
V_{soliton} &\rightarrow& - (1 + g^{+})\ln R + g^{+}\ln T,\nb\\
M_{soliton} &\rightarrow& - \frac{1}{2}g^{+}(2 + g^{+})\ln R \nb\\
& & + g^{+}(1 + g^{+})\ln T,
 \eqn
where  
\bq
\lb{3.42b}
g^{+} \equiv  2\sum^{(N+n)/2}_{k = 1 + n/2}{\alpha^{+}_{k}}.
\eq

\begin{figure}[htbp]
  \begin{center}
    \leavevmode
    \epsfig{file=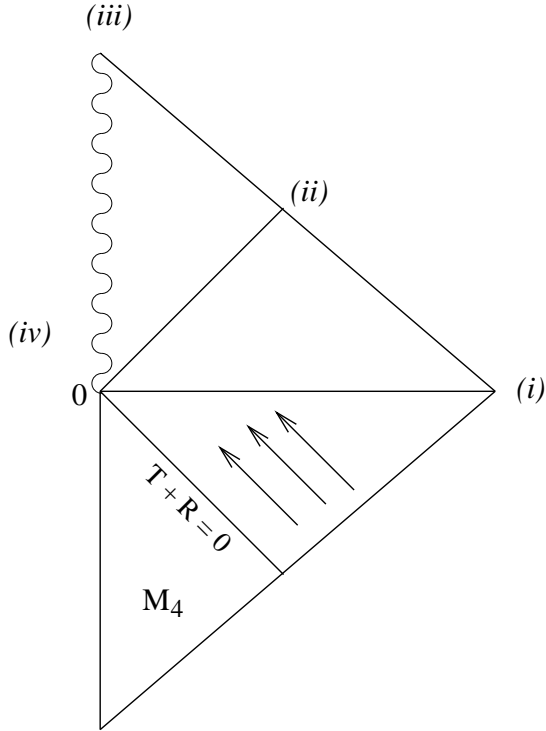,width=0.4\textwidth,angle=0}
\caption{The corresponding Penrose diagram for gravitational collapse of
purely gravitational cylindrical waves. The wave-front is denoted by the
straight line $T+R = 0$. before the waves arrives, the spacetime is flat,
denoted by $M_{4}$. The spacetime is asymptotically flat at spacelike infinite
$|T| \ll  R \rightarrow \infty$ in the radial direction. When the wave-front
arrives at the symmetry axis ($T, R ) = (0, 0$), in general part of the wave is
reflected to infinity along the light cone $T - R = 0$ and part of it is
focused to spacetime singularity on the axis, denoted by the curved vertical
line.}     
\label{fig3}   
\end{center} 
\end{figure}

Then, we find that the spacetime in this region is always singular,
except for the cases 
\bq
\lb{3.43}
(i) \;\; 0 \le g^{+} \le 1, \;\;\; {\mbox{or }} \;\;\; (ii) \;\; g^{+} = -2.
\eq
When $g^{+} = 0$, we can see that the collapsing gravitational waves are
reflected completely from the symmetry axis to spacelike infinity along the
future light cone, and nothing is left behind, so that the final state of the
spacetime is Minkowskian.

(iv) {\em The focusing region, $R \sim T  \rightarrow 0^{+}$}:  This
is the region where the collapsing gravitational waves just arrive at the axis,
and start to accumulate there. It can be shown that 
\bqn
\lb{3.44a}
\sigma_{k} &\rightarrow& 4\frac{\left(T^{0}_{k}\right)^{2}
+ w_{k}^{2}}{R^{2}},  \nb\\ 
\gamma_{k} &\rightarrow& \gamma^{0}_{k},
\eqn
and
\bqn
\lb{3.44b}
V_{soliton} &\rightarrow& - (1 + g^{+})\ln R, \nb\\
M_{soliton} &\rightarrow& - \frac{1}{2}g^{+}(2 + g^{+})\ln R,
 \eqn
where $\gamma^{0}_{k}$ is a constant. Thus, the spacetime in this region is
singular, unless the conditions given by Eq.(\ref{3.43}) hold.  

In review of all the above, we have the following: The only solutions that
represent gravitational collapse of purely gravitational cylindrical waves in
an otherwise flat spacetime background are given by Eq.(\ref{3.39}) subjected
to the conditions of Eqs.(\ref{3.39aa}) and (\ref{3.40b}). After the waves
arrive at the symmetry axis, part of them is reflected to spacelike infinity
along the light cone $ R \sim T \rightarrow \infty$, and part of them is
focused to  form spacetime singularity on the axis, except for the cases
defined by Eq.(\ref{3.43}).  The
corresponding Penrose diagram is given by Fig. 3.

\section{Conclusions}

In this paper, the general matching of two cylindrically symmetric regions
across a null hypersurface is studied using the Barrab\'es-Israel method. It is
shown that in general the hypersurface supports a null dust shell. As an
application of these formulae, we study the collapse of a matter field that
satisfies the condition $R_{AB}g^{AB} = 0, \; (A, B = z, \varphi)$, in an
otherwise flat spacetime background. It is found that the only case that
represents gravitational collapse is that where the normal vector to the
hypersurface $\Phi = Const.$ is space-like, where $\Phi \equiv
(g_{zz}g_{\varphi\varphi})^{1/2}$. As a result, the gravitational collapse of
a purely gravitational cylindrical  wave or a pure null dust fluid cannot be
realized in a flat spacetime background. 

The general formulae are further applied to the collapse of purely
gravitational cylindrical waves in a flat spacetime background  and the general
conditions for such collapse are found. It is shown that after the waves
arrive at the axis, part of them is reflected to spacelike infinity along
the future light cone, and part of it is focused to form   spacetime
singularities on the symmetry axis. The only  cases where no spacetime
singularity is formed are these given by Eq.(\ref{3.43}). 

\section*{Acknowledgment}

The financial assistance from CNPq  (AZW)  is gratefully acknowledged.

\end{document}